\documentclass[pra,showpacs,superscriptaddress,eqsecnum,twocolumn]{revtex4}
\usepackage{graphicx}
\usepackage{times}

\begin{document}

\title{Minimal qubit tomography}

\author{Jaroslav \v{R}eh\'{a}\v{c}ek}
\affiliation{Department of Optics, Palacky University, 17.\ listopadu 50, 
772~00 Olomouc, Czech Republic}

\author{Berthold-Georg Englert}
\affiliation{Department of Physics, National University of Singapore, 
Singapore 117542, Singapore}

\author{Dagomir Kaszlikowski}
\affiliation{Department of Physics, National University of Singapore, 
Singapore 117542, Singapore}

\date{15 May 2004}  

\begin{abstract}
We present, and analyze thoroughly, a highly symmetric and efficient scheme
for the determination of a single-qubit state, such as the polarization
properties of photons emitted by a single-photon source.
In our scheme there are only four measured probabilities, just enough for the
determination of the three parameters that specify the qubit state, whereas
the standard procedure would measure six probabilities. 
\end{abstract}


\maketitle

\section{Introduction}\label{sec:Intro}
Experiments that exploit the polarization degree of freedom of single photons,
detected one by one, have become an almost routine matter in recent years.
In particular, a whole class of experiments that demonstrate the technical
feasibility of quantum cryptography, or quantum key distribution, use the
photon polarization as the carrier of the quantum bit, or \emph{qubit}.
Other experiments make use of a spatial degree of freedom, essentially the
path qubit of a two-path interferometer, which is sometimes translated into the
alternative of early or late arrival for the sake of easier transmission.

In applications like these, as well as many others, one must be able to
characterize the qubit source and the transmission channel. 
For this purpose a complete determination of the state of the qubit is
required, both as it is emitted from the source and as it arrives after
transmission.
To be able to perform the regular on-the-fly calibration of the setup, so as
to compensate for the unavoidable drifts, one needs an efficient diagnostics
that does not consume more qubits than really necessary.

The standard procedure measures three orthogonal components of the relevant
qubit analog of Pauli's spin vector operator, so that \emph{six} probabilities
are estimated for the determination of the \emph{three} real parameters that
specify the qubit state. 
But clearly, \emph{four} measured probabilities should suffice to establish
the values of three parameters.
Indeed, such minimal schemes for state determination are possible, and it is
the objective of this paper to present and analyze one such scheme, a highly
symmetric one.

In Sec.~\ref{sec:QubitTomo} we briefly review the standard six-output
measurement scheme and then introduce the minimal four-output scheme, followed
by remarks on state determination for qubit pairs.
We then proceed to describe, in Sec.~\ref{sec:ElliMeters}, optical
implementations for the measurement of a photon's polarization qubit ---
polarimeters or ellipsometers in the jargon of classical optics.

The question of how one infers a reliable estimate for the qubit state after
the detection of a finite, possibly small, number of qubits is addressed in
Sec.~\ref{sec:Counting}.
After discussing the optimality of the highly symmetric four-output scheme in
Sec.~\ref{sec:optimal}, and remarking on some peculiar aspects of measuring
pure qubit states in Sec.~\ref{sec:pure}, we analyze adaptive measurement
strategies in Sec.~\ref{sec:adaptive}, and then close with a summary.

\section{Qubit tomography}\label{sec:QubitTomo}
\subsection{Standard six-state tomography}\label{sec:standard6}
We describe, as usual, the binary quantum alternative of the qubit by a Pauli
vector operator $\vec{\sigma}=(\sigma_x,\sigma_y,\sigma_z)$.
The physical nature of the qubit is irrelevant for the present general 
discussion 
--- it might just as well be the spin-$\frac{1}{2}$ degree of
freedom of an electron, or a pseudo-spin such as the path degree of freedom 
in a two-path interferometer or the internal degree of freedom of a two-level
atom --- 
but in the particular application that we have in mind it is the polarization
degree of freedom of a photon. 
Then we use the convention specified by
\begin{eqnarray}
  \label{eq:A1}
  \sigma_x&=&\pketbra{h}{v}+\pketbra{v}{h}\,,\nonumber\\
  \sigma_y&=&i\left(\pketbra{h}{v}-\pketbra{v}{h}\right)\,,\nonumber\\
  \sigma_z&=&\pketbra{v}{v}-\pketbra{h}{h}\,,
\end{eqnarray}
where \pket{v} and \pket{h} are the ket vectors for
vertical and horizontal polarization, respectively.

The statistical operator of the qubit emitted by a given source,
\begin{equation}
  \label{eq:A2}
  \rho=\frac{1}{2}\bigl(1+\vec{s}\cdot\vec{\sigma}\bigr)\,,
\end{equation}
is parameterized by the Pauli vector 
$\vec{s}=\expect{\vec{\sigma}}=\tr{\vec{\sigma}\rho}$, the expectation
value of $\vec{\sigma}$. 
The positivity of $\rho$ restricts the Pauli vectors to the Bloch ball,
$s=\bigl|\vec{s}\bigr|\leq1$.
The experimental characterization of the source requires, therefore, 
a complete measurement of $\vec{s}$ with sufficient precision.
Any procedure that can yield this information is an example of 
\emph{qubit tomography}.

In the standard approach one measures $\sigma_x$ for some qubits supplied by
the source, $\sigma_y$ for some others, and $\sigma_z$ for yet others.
Assuming an unbiased procedure, that is for each qubit there is an equal
chance for either one of the three measurements to happen, there are six
possible outcomes that occur with the probabilities 
\begin{equation}
  \label{eq:A3}
  p_{\xi\pm}=\expect{\frac{1}{6}(1\pm\sigma_\xi)}\equiv\expect{P_{\xi\pm}} 
\quad\text{for}\enskip  \xi=x,y,z\,.
\end{equation}
Each operator $P_{\xi\pm}$ is a third of a projector, and since these 
nonnegative operators decompose the identity,
\begin{equation}
  \label{eq:A4}
  \sum_{\xi=x,y,z}\left(P_{\xi+}+P_{\xi-}\right)=1\,,
\end{equation}
they constitute the Positive Operator Valued
Measure (POVM) of this standard \emph{six-state tomography}.

This POVM is an example of a tomographically complete set of measurements of
pairwise complementary observables, namely $\sigma_x$, $\sigma_y$, and
$\sigma_z$, so that their eigenstates constitute sets of mutually unbiased
bases. 
As Wootters and Fields have shown \cite{WooFie}, such sets are particularly
well suited for tomographic purposes, inasmuch as the statistical errors in
the estimates based on a finite number of measurements are minimal.
The sets themselves are not of minimal size, however, because one measures six
probabilities to determine three parameters, the components of the Pauli
vector $\vec{s}$. 
Indeed, the six probabilities of \Eqref{A3} are subject to the three
constraints $p_{\xi+}+p_{\xi-}=\frac{1}{3}$, $\xi=x,y,z$, rather than to the
single constraint of unit sum.
A minimal POVM, by contrast, would refer to only four outcomes and their
probabilities, with unit sum as the only constraint.

\subsection{Minimal four-state tomography}\label{sec:minimal4}
\begin{figure}[t]
\centerline{\includegraphics{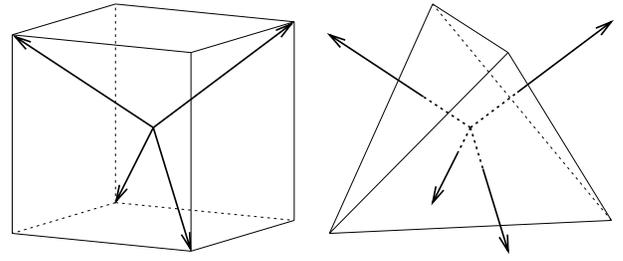}}
\caption{\label{fig:quartet}%
Picturing the vector quartet of \Eqref{B1} as pointing from the center to four
non-adjacent corners of a cube, 
or as the vectors normal on the faces of a tetrahedron.}
\end{figure}

We construct such a minimal POVM of high internal symmetry by first choosing
four unit vectors, $\vec{a}_1$, \dots, $\vec{a}_4$, with equal angle between
each pair,
\begin{equation}
  \label{eq:B1}
  \vec{a}_j\cdot\vec{a}_k=\frac{4}{3}\delta_{jk}-\frac{1}{3}
=\left\{
  \begin{array}{r@{\ \text{for}\ }l}
  1 & j=k\,,\\ -1/3 & j\neq k\,.
  \end{array}
\right.
\end{equation}
Geometrically speaking, such a quartet consists of the vectors pointing from
the center of a cube to non-adjacent corners, as illustrated in
Fig.~\ref{fig:quartet} and exemplified by
\begin{eqnarray}
  \label{eq:B2}
  \vec{a}_1&=&3^{-1/2}(1,1,1)\,,\nonumber\\
  \vec{a}_2&=&3^{-1/2}(1,-1,-1)\,,\nonumber\\
  \vec{a}_3&=&3^{-1/2}(-1,1,-1)\,,\nonumber\\
  \vec{a}_4&=&3^{-1/2}(-1,-1,1)\,.
\end{eqnarray}
Alternatively, one may picture these vectors as the normal vectors for the
faces of the tetrahedron that is defined by the other four corner of the cube.

The linear dependence of the $\vec{a}_j$'s is stated by their null sum, 
\begin{equation}
  \label{eq:B3}
  \sum_{j=1}^4\vec{a}_j=0\,,
\end{equation}
and their completeness by the decomposition of the unit dyadic,
\begin{equation}
  \label{eq:B4}
  \frac{3}{4}\sum_{j=1}^4\vec{a}_j\vec{a}_j=\tensor{1}\,.
\end{equation}
The perfect symmetry of the tetrahedron geometry manifests itself in the
simplicity of this completeness relation and the inner products of \Eqref{B1}.
As discussed in Sec.~\ref{sec:tetra-opt} below, the tetrahedron geometry is 
optimal in the sense that any other vector quartet would define a less 
efficient scheme for four-state tomography. 

Each such quartet of $\vec{a}_j$'s defines  
a POVM for minimal four-state tomography in accordance with
\begin{equation}
  \label{eq:B5}
\sum_{j=1}^4P_j=1\quad\text{with}\enskip  
P_j\equiv\frac{1}{4}\left(1+\vec{a}_j\cdot\vec{\sigma}\right)\,.
\end{equation}
This POVM is an example of a ``symmetric informationally complete POVM'' 
\cite{Renes+3:03}. 
Upon measuring it and so determining the probabilities \cite{F+W+GT}
\begin{equation}
  \label{eq:B6}
  p_j=\expect{P_j}=\frac{1}{4}\left(1+\vec{a}_j\cdot\vec{s}\,\right)\,,
\end{equation}
the Pauli vector is readily available,
\begin{equation}
  \label{eq:B7}
  \vec{s}=3\sum_jp_j\vec{a}_j\,,
\end{equation}
and so are the statistical operator and its square,
\begin{eqnarray}
  \label{eq:B8}
  \rho&=&6\sum_jp_jP_j -1=\sum_j\expect{P_j}\left(6P_j-1\right)\,,\nonumber\\
  \rho^2&=&\rho-1+3\sum_jp_j^2\,.
\end{eqnarray}
It follows that, in addition to being restricted to the range 
$0\leq p_j\leq\frac{1}{2}$, the probabilities $p_j$ obey the inequalities
\begin{equation}
  \label{eq:B9}
  \frac{1}{4}\leq\sum_jp_j^2=\frac{3+s^2}{12}\leq\frac{1}{3}\,.
\end{equation}
The upper bound is reached by all pure states, $\rho=\rho^2$ and $s=1$, 
the lower bound for the completely mixed state, $\rho=\frac{1}{2}$ and $s=0$.

\subsection{Qubit-pair tomography}\label{sec:2qubits}
The minimal property of the four-state POVM of \Eqref{B5} carries over to
multi-qubit states. 
In case of $n$ qubits, one has $4^n$ joint probabilities for the $4^n-1$
independent parameters of the $2^n\times2^n$ matrix elements of the
statistical operator, so that the count is just right.
By contrast, if one were to measure $n$ realizations of the six-state POVM of
\Eqref{A4}, one would have $6^n$ joint probabilities which contain quite a lot
of redundant information.

More specifically, consider the ${n=2}$ situation of a source emitting qubit
pairs.  
Using the $P_j$'s from above for one qubit and corresponding
operators $Q_k$ for the other, we obtain the 16 joint probabilities
\expect{P_jQ_k} by measuring the two four-state POVMs.
They are the numerical ingredients in 
\begin{equation}
  \label{eq:C1}
  \rho=\sum_{j,k}\left(6P_j-1\right)\expect{P_jQ_k}\left(6Q_k-1\right)\,,
\end{equation}
the explicit construction of the two-qubit statistical operator. 
If there are no correlations in the joint probabilities, so that
\expect{P_jQ_k}=\expect{P_j}\expect{Q_k}, then this $\rho$ is the product of
two factors of the one-qubit form in \Eqref{B8}, as it should be.
The generalization of the ${n=2}$ expression \refeq{C1} to ${n>2}$ is
immediate. 

Qubit-pair tomography of this kind requires that the vector quartet
$\vec{b}_k$ associated with the $Q_k$'s has a known orientation relative to
the quartet $\vec{a}_j$ of the $P_j$s.
One can determine this orientation by 
``quantum measurement tomography'' \cite{pedantry}, that is 
by measuring the joint probabilities $\expect{P_jQ_k}$ for  
a source with a known output \cite{ChK}.
In the simplest situation, for example, the source emits pairs that are
perfectly anticorrelated,
\begin{equation}
  \label{eq:C2}
  \rho=\frac{1}{4}\left(1-\vec{\sigma}^{(1)}\cdot\vec{\sigma}^{(2)}\right)\,.
\end{equation}
The orthogonal dyadic $\tensor{O}$ that turns the $\vec{a}_j$ quartet and the
$\vec{b}_k$ quartet into each other,
\begin{equation}
  \label{eq:C3}
  \vec{b}_k=\tensor{O}\cdot\vec{a}_k\,,\qquad
  \vec{a}_j=\vec{b}_j\cdot\tensor{O}\,,
\end{equation}
is then given by
\begin{equation}
  \label{eq:C4}
  \tensor{O}=-9\sum_{j,k}\vec{a}_j\expect{P_jQ_k}\vec{a}_k
              =-9\sum_{j,k}\vec{b}_j\expect{P_jQ_k}\vec{b}_k\,,
\end{equation}
and can thus be determined experimentally.

\section{Ellipsometry}\label{sec:ElliMeters}
\subsection{Standard six-outcome ellipsometer}
\begin{figure}[t]
\centerline{
\includegraphics{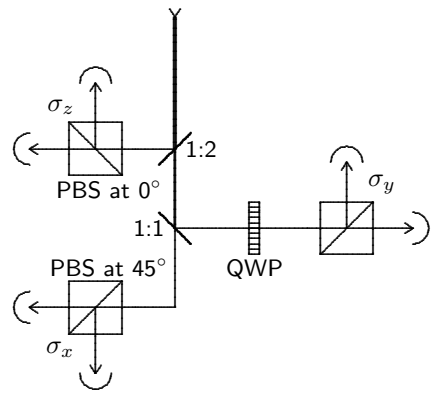}
}
\caption{ \label{fig:standard}%
Schematic setup of an ellipsometer that implements the standard six-state
POVM; see text. 
}
\end{figure}

Devices for characterizing the polarization properties of a light source are
called \emph{ellipsometers}, a term that makes reference to elliptic
polarization as the generic outcome of the measurement. 
Figure \ref{fig:standard} shows the schematic setup of a standard six-state
ellipsometer, which realizes the POVM of Eqs.~\refeq{A3} and \refeq{A4} for 
the photonic polarization qubit.
The input beam splitter (BS) reflects one third of the light to a polarizing
BS (PBS) that reflects vertically polarized photons, and transmits horizontally
polarized ones, and so directs them to two photodetectors. 
This branch thus realizes a measurement of $\sigma_z$ and accounts for
$P_{z+}$ and $P_{z-}$ in the sum of \Eqref{A4}.

Two thirds of the light are transmitted at the input BS and are then equally
split at a second BS. 
A photon transmitted there will be detected by either one of the two detectors
behind another PBS. 
This PBS is rotated by $45^\circ$, so that a measurement of $\sigma_x$ 
is realized in this branch and the term
$P_{x+}+P_{x-}$ is accounted for in \Eqref{A4}. 

Finally, the photons that are reflected at the second BS pass through a
quarter-wave plate (QWP) before a PBS directs them to a third pair of
detectors.
This branch implements the measurement of $\sigma_y$ and accounts for the
remaining terms in \Eqref{A4}, namely $P_{y+}$ and $P_{y-}$.

\subsection{Minimal four-outcome ellipsometer}
There are a number of alternative schemes for an ellipsometer that realizes 
the minimal four-state POVM of \Eqref{B5}. 
To demonstrate the case, we present one scheme here and discuss a few other
schemes, which are much simpler and much more practical, 
elsewhere \cite{1loop,newSetup}.
 
\begin{figure}[t]
\centerline{
\includegraphics{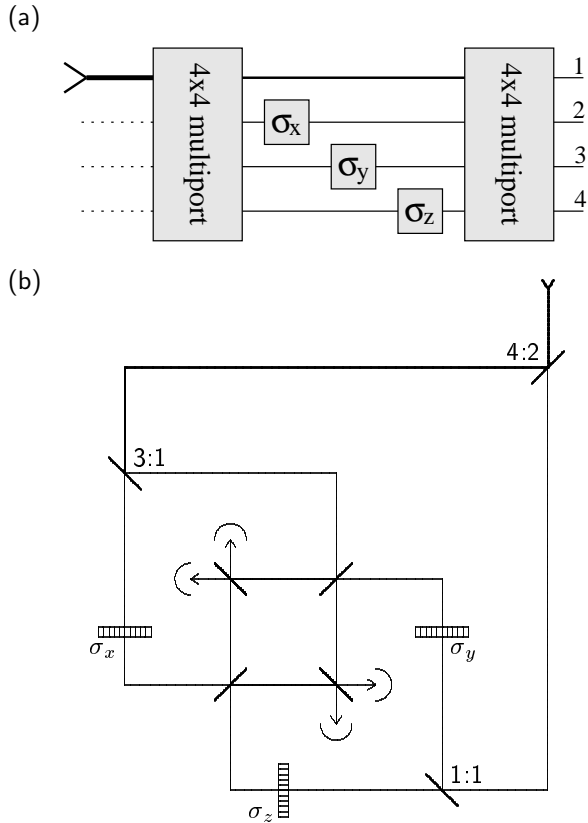}
}
\caption{ \label{fig:minimal}%
(a)~Principle of a four-path interferometer for minimal ellipsometry;
(b)~Optical network for the implementation;
see text.}
\end{figure}

The principle of one minimal ellipsometer is illustrated by
Fig.~\ref{fig:minimal}(a).
It is an asymmetric four-path interferometer.
At the input, one half of the light intensity is directed into the uppermost 
path for reference, whereas each of the other three paths gets one sixth of
the intensity. 
Photons in the lower paths pass through wave plates that realize the unitary
polarization transformations $\rho\to\sigma_\xi\rho\sigma_\xi$ with $\xi=x$, or
$y$, or $z$, respectively.
Then all four paths are recombined by a balanced beam merger, which
distributes the intensity of each input evenly among the four outputs.
An unpolarized photon has, therefore, a probability of 25\% for being detected
by a particular one of the four detectors. 
The probabilities for a polarized photon are the $p_j$'s of \Eqref{B6}, so
that the polarization POVM of \Eqref{B5} is measured indeed.

An optical network for a four-path interferometer of this kind is shown in
Fig.~\ref{fig:minimal}(b). 
The BS at the input reflects $4/6$ of the intensity and transmits $2/6$,
and the subsequent BSs either split the beam $3:1$ or $1:1$;
together these three BSs implement the initial stage of
Fig.~\ref{fig:minimal}(a).   
At the central stage, there are wave plates in three partial beams (labeled by 
$\sigma_x$, $\sigma_y$, and $\sigma_z$, respectively).
And the final beam merger consists of four 1:1 BSs which direct the photons to
the four photodetectors \cite{multiport}.

The interferometer of Fig.~\ref{fig:minimal}(b) has several loops.
Some alternative setups have two loops \cite{RenesDiss}, 
or need a single loop only, among them the interferometer
of the experiment by Clarke \textit{et al.} \cite{Clarke}
and the minimal one-loop scheme of Ref.~\cite{1loop}, and yet another
setup has no loop at all \cite{newSetup}.
We note that Clarke \textit{et al.\/} did not perform ellipsometry, their
experiment served a different 
purpose and, although their setup could be used for ellipsometry,
there is no mentioning of this possible application in Ref.~\cite{Clarke}.  

\begin{figure}[t]
\centering
\includegraphics{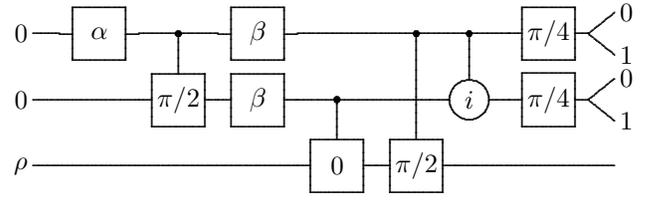}
\caption{ \label{fig:QCnet}%
Quantum computation network for minimal qubit tomography.
The bottom qubit is the one of interest, it enters in the state $\rho$ whose
properties are to be determined.
The two top qubits are auxiliary qubits that enter in state $0$ of the
computational basis and are eventually measured.
The probabilities for the four different measurement results --- $00$, $01$,
$10$, and $11$ --- are the $p_j$'s of \Eqref{B6} with the $\vec{a}_j$'s of
\Eqref{B2}.    
The network uses eight gates of the generalized Hadamard type \refeq{C5}, 
three of them controlled by one
of the auxiliary qubits, and a controlled phase gate.}
\end{figure}

It is worth mentioning that such a setup can also be viewed as a quantum
computation network for three qubits, one being the polarization qubit
of interest, the other two qubits representing the four paths of the 
interferometer \cite{1aux}. 
The network is depicted in Fig.~\ref{fig:QCnet}.
It consists of a sequence of generalized Hadamard gates,
\begin{equation}
  \label{eq:C5}
\frac{\quad}{}\!\fbox{$\phi$}\!\frac{\quad}{}\,:\quad
\left(  \begin{array}{c} \bra{0} \\[1ex] \bra{1}\end{array}\right)\to
\left(  \begin{array}{cc} \cos\phi & \sin\phi\\ \sin\phi & -\cos\phi
\end{array}\right)
\left(  \begin{array}{c} \bra{0} \\[1ex] \bra{1}\end{array}\right)\,.
\end{equation}
At the first stage, the auxiliary qubits are prepared in a superposition state
that has amplitude $1/\sqrt{2}$ for $\ket{00}\widehat{=}\pket{vv}$ and 
amplitudes $1/\sqrt{6}$ for $\ket{01}$, $\ket{10}$, 
and $\ket{11}\widehat{=}\pket{hh}$. 
We achieve this by a controlled gate with $\phi=\frac{1}{2}\pi$ (this is a
controlled-not gate, in fact) that is sandwiched by a gate with $\phi=\alpha$ 
and two gates with $\phi=\beta$, where $\alpha$ and $\beta$ are such that
$\sin(2\alpha)=\frac{1}{3}(\sqrt{3}-1)$ and $\tan(2\beta)=\sqrt{3}+1$.

The central stage has controlled gates acting on the qubit of interest,
a gate with $\phi=0$ for $\sigma_z$, another one with $\phi=\frac{1}{2}\pi$
for $\sigma_x$. For their product to realize $\sigma_y=i\sigma_x\sigma_z$,
we provide the factor of $i$ by a subsequent controlled phase gate, which 
implements the phase change $\ket{11}\to i\ket{11}$, but has no effect
on $\ket{00}$, $\ket{01}$, and $\ket{10}$.

At the final stage, the two auxiliary qubits are passed through standard
Hadamard gates, for which $\phi=\frac{1}{4}\pi$ in \Eqref{C5}, and are then
measured in the computational basis.  
The probabilities for getting $00$, $01$, $10$, or $11$ are $p_1$, $p_4$,
$p_2$, and $p_3$ of Eqs.~\refeq{B6} and \refeq{B2}, respectively. 

Conditioned on the measurement outcome, the qubit of interest emerges in the
corresponding reduced state $2P_j\rho P_j/p_j$.
For the optical implementation of Fig.~\ref{fig:minimal}(b), 
these final states of the polarization qubit are fictitious, however, 
unless the photodetectors are of a fantastic non-demolition kind:
sensitive to the passage of a photon without absorbing the photon or
affecting its polarization~\cite{1new-1aux}.

\section{Counting qubits}\label{sec:Counting}
\subsection{Maximum-likelihood estimator}\label{sec:MLest}
In an actual experiment, we do not measure the probabilities of Eqs.\
\refeq{A3} or \refeq{B6}, but rather relative frequencies that are
statistically determined by these probabilities. 
The available information consists of the counts of detector clicks, 
$n_1$, $n_2$, $n_3$, $n_4$ for the minimal four-state tomography of
Sec.~\ref{sec:minimal4}
and $n_{x\pm}$, $n_{y\pm}$, $n_{z\pm}$ for the standard six-state scheme of
Sec.~\ref{sec:standard6}.
In what follows, we focus on the novel four-detector situation.

In view of the intrinsic probabilistic nature of quantum phenomena, a given
total number of $N$ qubits does not result in a definite, predictable number
of clicks for each detector. 
It is, therefore, clear that an observed break up, 
\begin{equation}
  \label{eq:D1}
  N=n_1+n_2+n_3+n_4\,,
\end{equation}
is consistent with not just one Pauli vector $\vec{s}$ in \Eqref{A2}, but
with many.
One expects that, as a rule, the obvious guess that obtains from \Eqref{B7}
upon the replacement $p_j\to n_j/N$, namely
\begin{equation}
  \label{eq:D2}
    \vec{S}=3\sum_j\nu_j\vec{a}_j 
\qquad\text{with}\enskip\nu_j\equiv\frac{n_j}{N}\,,
\end{equation}
gives a reliable estimate of the true Pauli vector $\vec{s}$.
But it may happen that the length of this inferred $\vec{S}$ exceeds unity, 
and this is in fact a typical situation if the true $\vec{s}$ is of unit
length, that is the source emits qubits in a pure state.

Let us thus proceed to show how one infers  
a plausible and physically correct answer on the basis of the registered data. 
The experimental data consists in fact not just of the total counts $n_j$ 
in the break up \refeq{D1}, but of a particular sequence of detector clicks. 
\emph{Provided that} 
the source emits qubits in the state specified by the Pauli vector $\vec{s}$,
with no statistical correlations between different qubits, 
the probability $\mathcal{L}(\vec{s};n_1,\dots,n_4)$ 
for getting the observed click sequence is, therefore, given by
\begin{equation}\label{eq:D3}
\mathcal{L}=\prod_{j=1}^4 {p_j(\vec{s})}^{n_j}\,,
\end{equation}
where $p_j(\vec{s})$ is the probability \refeq{B6} that a qubit is registered
by the $j$th detector.

\begin{figure}[t]
\centerline{\includegraphics{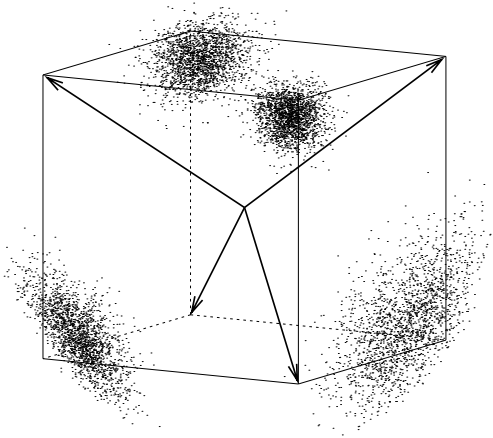}}  
  \caption{\label{fig:clouds}%
The likelihood function as a probability density for the Pauli vector
$\vec{s}$.
The plot shows the ``likelihood clouds'' for simulated measurements for which
the most likely Pauli vector has length $s=0.84$ and is pointing toward one of
the corners of the cube of Fig.~\ref{fig:quartet}.
A total of $100$ qubits have been detected for the bottom right corner,
$200$ for the bottom left corner, $400$ for the top rear corner, and $800$ for
the top front corner. 
The successive shrinking of the cloud is clearly visible. 
}
\end{figure}

Conversely, in the spirit of the Bayesian principle of statistical inference, 
we can regard $\mathcal{L}(\vec{s};n_1,\dots,n_4)$ as the
likelihood that the source is characterized by $\vec{s}$, \emph{given that} 
a click sequence with total detector counts $n_1$, \dots, $n_4$ was observed.
When many qubits have been counted, the likelihood function is sharply peaked
and essentially vanishes outside the immediate vicinity of its maximum.
These matters are illustrated by the four ``likelihood clouds'' in
Fig.~\ref{fig:clouds}. 
 
The maximum-likelihood (ML) estimator $\vec{S}$ picks 
out the most likely Pauli vector, the one for which  $\mathcal{L}$ is
largest,
\begin{equation}
  \label{eq:D4}
  \max_{\mbox{\footnotesize$\vec{s}$}}\mathcal{L}(\vec{s})
=\mathcal{L}(\vec{S})\,.
\end{equation}
For the purposes of this paper, 
we accept this $\vec{S}$ as our plausible guess for $\vec{s}$, while being
fully aware of other strategies \cite{BayesAver1,Helstrom,LNP}.

An important theorem by Fisher states that ML estimators
become efficient in the large-$N$ limit \cite{Fisher,cramer}. 
So the performance, for large $N$, of an experimental tomographic setup can 
be quantified by the accuracy of the ML estimator. 
Usually an analytical expression for the ML estimator is not available and one
has to solve a nonlinear operator equation for
$\rho=\frac{1}{2}(1+\vec{S}\cdot\vec{\sigma})$ to find
it~\cite{zdenek}. 
In the present context, however, the high symmetry of the vector quartet 
of \Eqref{B1} makes it possible to simplify this problem considerably. 
 
There is a benefit in maximizing the right-hand side of  
\Eqref{D2} with respect to the probabilities $p_j$ rather than with respect
to $\vec{s}$. 
Lagrange multipliers are used to account for the two constraints. 
One is the unit sum of the $p_j$, $\sum_j p_j=1$, that is the unit trace of
$\rho$, the other is the positivity of $\rho$, $\rho\geq0$, which is the
upper bound in \Eqref{B9}. 

Without this positivity constraint, the likelihood \refeq{D3} would be
maximized by $p_j=\nu_j$, which in turn would imply the Pauli vector of
\refeq{D2}.  
But, if the actual counts violate the inequality 
\begin{equation}\label{eq:D5}
\sum_j \nu_j^2\leq \frac{1}{3}\,,
\end{equation}
this simple estimation fails to provide a physically meaningful result.
When this happens, both constraints must be taken into 
account and the likelihood $\mathcal{L}(\vec{s})$ attains its maximum
on the boundary of the set of qubit states, that is for a Pauli vector of
unit length \cite{BayesAver2}.  

The variation of the likelihood vanishes at the extremal point,
\begin{equation}\label{eq:D6}
\delta\log\mathcal{L}=\sum_j\frac{n_j}{p_j}\delta p_j=0\,.
\end{equation}
The variations $\delta p_j$ are subject to the two constraints
\begin{equation}
  \label{eq:D7}
  \sum_j\delta p_j=0\,,\qquad \sum_jp_j\delta p_j=0\,,
\end{equation}
for which we use the Lagrange multipliers ${N\lambda}$ and ${3N\mu}$, 
respectively. 
Denoting the $p_j$ values at the extremal point by $\tilde{p}_j$,
\refeq{D6} then implies
\begin{equation}\label{eq:D8}
\frac{\nu_j}{\tilde{p}_j}=\lambda+3\mu \tilde{p}_j \qquad\text{for}
\quad j=1,\dots,4\,.
\end{equation}
We exploit $\sum_j\nu_j=1$, $\sum_j \tilde{p}_j=1$, and
$\sum_j\tilde{p}_j^2=\frac{1}{3}$ to establish 
\begin{equation}\label{eq:D9}
\sum_j\frac{\nu_j}{\tilde{p}_j}=4\lambda+3\mu
\end{equation}
and
\begin{equation}\label{eq:D10}
1=\lambda+\mu\,,
\end{equation}
the first by summing the four equations in \refeq{D8}, the second by summing
them after multiplication with $\tilde{p}_j$.

Taken together, Eqs.~\refeq{D8}--\refeq{D10} make up a
set of six equations for the six unknowns: $\tilde{p}_1$, \dots, 
$\tilde{p}_4$, $\lambda$, and $\mu$.
We solve the quadratic equations \refeq{D8} for $\tilde{p}_j$, 
or rather $\nu_j/\tilde{p}_j$,
\begin{equation}
  \label{eq:D11}
  \frac{\nu_j}{\tilde{p}_j}
=\frac{1}{2}\left(\lambda+\sqrt{\lambda^2+12\mu \nu_j}\right)\,,
\end{equation}
and \refeq{D10}
for $\lambda$ and substitute these into \Eqref{D9} to arrive at
a single equation for $\mu$,
\begin{equation}\label{eq:D12}
\mu+2-\frac{1}{2}\sum_j\sqrt{(1-\mu)^2+12\mu \nu_j}=0\,.
\end{equation}
There is always the solution $\mu=0$, and thus $\lambda=1$, but \refeq{D11}
amounts to \refeq{D2} for these values, and therefore this solution is not
acceptable, when the inequality \refeq{D5} is not obeyed, as is the case in
the present discussion. 
Upon recognizing that  \Eqref{D12} can also be written as
\begin{equation}
  \label{eq:D13}
  6\mu^2\sum_j
\frac{(3\nu_j-1)\nu_j}{1-\mu+6\mu \nu_j+\sqrt{(1-\mu)^2+12\mu \nu_j}}=0
\end{equation}
we can discard the unphysical solution $\mu=0$ and find the relevant solution
in the range $0<\mu\leq2$ as the root of this sum over $j$.
The upper bound $\mu=2$ is reached when all qubits are detected by the same
detector, so that one of the $\nu_j$ equals $1$ and the others vanish.

In summary, we determine the ML estimator 
\begin{equation}
  \label{eq:D14}
  \vec{S}=3\sum_j\tilde{p}_j\vec{a}_j
\end{equation}
as follows. 
If the inequality \refeq{D5} is obeyed by $\nu_j=n_j/N$, we take 
$\tilde{p}_j=\nu_j$.
Otherwise we find $\mu$ as the positive root of \refeq{D12} or \refeq{D13} and
then get the four $\tilde{p}_j$'s from \refeq{D11} with $\lambda=1-\mu$.
In the extremal situation of $\nu_j=\delta_{jk}$, we have
$\tilde{p}_j=\frac{1}{6}+\frac{1}{3}\delta_{jk}$ and $\vec{S}=\vec{a}_k$.

\subsection{Many detector clicks}\label{sec:ManyClicks}
In this procedure for finding the ML estimator, there is a crucial difference
between relative frequencies $\nu_j$ that obey the inequality \refeq{D5}, and
can therefore serve as probabilities, and those that violate the inequality.
When the total number $N$ of detected qubits is small, statistical
fluctuations are relatively large and a violation is hardly surprising. 
But what is the typical situation for a large number of detector clicks,
should we expect \refeq{D5} to be obeyed or violated?

For the likelihood \refeq{D3}, only the break-up \refeq{D1} matters, not the
particular sequence of detector clicks.  
There are $N!/(n_1!n_2!n_3!n_4!)$ different sequences for a given break-up,
so that there is a multinomial statistics for the probability of getting a
particular break-up for the given probabilities $p_j(\vec{s})$ of \Eqref{B6}.
We denote the corresponding averages over possible break-ups by over-bars, as
illustrated by
\begin{equation}
  \label{eq:E1}
  \overline{n_j}=N\overline{\nu_j}=Np_j
\end{equation}
and
\begin{equation}
  \label{eq:E2}
  \overline{\nu_j\nu_k}=p_jp_k+\frac{1}{N}\bigl(\delta_{jk}p_k-p_jp_k\bigr)\,.
\end{equation}
Upon recalling that $\sum_jp_j^2=(3+s^2)/12$, cf.\ \Eqref{B9}, 
the latter averages imply
\begin{eqnarray}
  \label{eq:E3}
  \sum_j\overline{\nu_j^2}&=&\frac{1}{N}+\frac{N-1}{N}\sum_jp_j^2
\nonumber\\
&=&\frac{1}{3}-\frac{1-s^2}{12}
 +\frac{9-s^2}{12N}\,,
\end{eqnarray}
so that, \emph{on average}, the inequality \refeq{D5} is violated for pure
states (for which $s=1$), and is obeyed for mixed states (for which $s<1$).
We thus expect that the detector counts for a pure qubit state will typically
violate inequality \refeq{D5}, whereas the counts for a mixed state will tend
to obey it.

As a more precise statement about this matter, we note that the fraction of
detector click sequences that violate the inequality is 
\begin{equation}
  \label{eq:E4}
  \text{prob(violation)}=\frac{1}{2}
  +\frac{1}{2}\mathrm{erf}\biggl(\frac{\sqrt{N}}{2\kappa}\Bigl(
      \sum_j\overline{\nu_j^2}-\frac{1}{3}\Bigr)\biggr)\,,
\end{equation}
where $\mathrm{erf}(\ )$ is the standard error function 
and $\kappa$ is given by
\begin{eqnarray}
  \label{eq:E5}
  \kappa^2&=&\sum_{j,k}p_jp_k(p_j-p_k)^2
\nonumber\\
    &=&2\sum_jp_j^3-2\Bigl(\sum_jp_j^2\Bigr)^2\,.
\end{eqnarray}
Equation \refeq{E4} applies for ${N\gg1}$; in order to derive it, first observe
that the central limit theorem ensures that 
\begin{eqnarray}
  \label{eq:E6}
&&\overline{\exp\Bigl(i\sum_j\alpha_j(\nu_j-p_j)\Bigr)}
\nonumber\\&=&
\exp\Bigl(-\frac{1}{2N}
\sum_{j,k}\alpha_j(\delta_{jk}p_k-p_jp_k)\alpha_k\Bigr)  
\end{eqnarray}
for large $N$ \cite{Braunstein}.
Then use this generating function for the mean values of products of the 
$\nu_j$'s to calculate the probability that 
$\sum_j\nu_j^2>\frac{1}{3}$, with consistent approximations for ${N\gg1}$.
  
For pure states, the argument of the error function in \refeq{E4} is 
$1/(3\kappa\sqrt{N})$, which is always positive, but decreases 
with growing $N$, so that click sequences that violate the inequality are 
more frequent than the ones obeying it, but they are not much more frequent.
A remarkable exception is the situation of the Pauli vector $\vec{s}$ being
exactly opposite to one of the directions of the vector quartet of
Eqs.~\refeq{B1} and \refeq{B2}.
Then one of the $p_j$'s vanishes and the other three are all equal to
$\frac{1}{3}$, so that $\kappa=0$ and there is a unit probability for 
getting a violation.

For mixed states, the argument of the error function is negative and increases
in magnitude $\propto\sqrt{N}$ (more about this in Sec.~\ref{sec:orient}). 
Accordingly, a violation of the inequality is highly improbable. 
In the extreme situation of the completely mixed state, $\rho=\frac{1}{2}$,
all $p_j$'s are $\frac{1}{4}$, so that $\kappa=0$ here too, and the
probability for violating the inequality vanishes.

\subsection{Asymptotic efficiency}\label{sec:asymptotics}
After detecting many qubits, we expect the ML estimator $\vec{S}$ to be quite
close to the true Pauli vector $\vec{s}$.
If inequality \refeq{D5} is obeyed, the average squared distance
\begin{eqnarray}
  \label{eq:E7}
  (\Delta\vec{s})^2&=&\overline{\bigl(\vec{S}-\vec{s}\bigr)^2}=
  \overline{\Bigl(3\sum_j(\nu_j-p_j)\vec{a}_j\Bigr)^2}
\nonumber\\
&=&\frac{12}{N}\Bigl(1-\sum_jp_j^2\Bigr)=\frac{9-s^2}{N}
\end{eqnarray}
is immediately available as a consequence of \Eqref{E2}.

The analysis is more involved when \refeq{D5} is not obeyed.
Let us consider once more the extreme situation of $\vec{s}$ being exactly 
opposite to one of the $\vec{a}_j$, such as $\vec{s}=-\vec{a}_4$, say, so that 
$p_4=0$ and $\nu_4=0$, and $\nu_1$, $\nu_2$, $\nu_3$ differ little from
$p_1=p_2=p_3=\frac{1}{3}$. 
To leading order in $\nu_j-\frac{1}{3}$, the solution of \Eqref{D12} is then
\begin{equation}
  \label{eq:E8a}
  \mu=1+\frac{9}{8}\sum_{j=1}^3\Bigl(\nu_j-\frac{1}{3}\Bigr)^2
\end{equation}
and the resulting ML estimator is
\begin{equation}
  \label{eq:E8b}
  \vec{S}=\frac{1}{2}\sum_{j=1}^3(1+3\nu_j)\vec{a}_j
         =\vec{s}+\frac{3}{2}\sum_{j=1}^3
                 \Bigl(\nu_j-\frac{1}{3}\Bigr)\vec{a}_j\,.
\end{equation}
In conjunction with \Eqref{E2}, we thus get the large-$N$ approximation 
\begin{equation}
  \label{eq:E8}
 (\Delta\vec{s})^2 =3\,\overline{\,\nu_1^2+\nu_2^2+\nu_3^2\,}-1=\frac{2}{N}
\end{equation}
in this case of perfect anti-alignment between $\vec{s}$ and the quartet of
$\vec{a}_j$'s. 
The sensitivity to small misalignments is discussed in Sec.~\ref{sec:NumSim}.

\begin{figure}[t]
\centerline{
\includegraphics{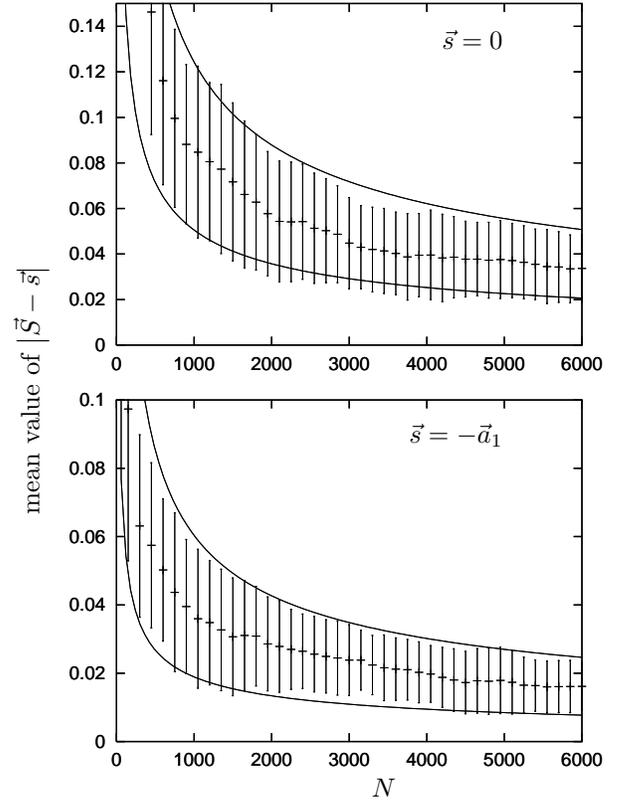}
}
\caption{\label{fig:MLasym}%
Mean value of the distance from the ML estimator $\vec{S}$ to the true Pauli
vector $\vec{s}$ for $40$ simulated experiments with up to $6000$ detected
qubits per run.
The top plot is for $\vec{s}=0$, the bottom plot for $\vec{s}=-\vec{a}_4$.
The solid lines indicate one standard deviation 
to each side of the mean value
, as obtained from the large-$N$
approximations for $(\Delta\vec{s})^2$ in Eqs.~\refeq{E7} and \refeq{E8}, 
respectively.
}
\end{figure}

As demonstrated by the numerical results of Fig.~\ref{fig:MLasym}, these
asymptotic approximations are actually quite reliable for $N\gtrsim1000$. 
The plots report the mean values, with statistical error bars of one standard
deviation, of the distance $\bigl|\vec{S}-\vec{s}\bigr|$ as obtained from $40$ 
simulated experiments, whereby up to $6000$ qubits are detected in each run.
The top plot refers to a true state that is completely mixed, so that $s=0$ in
\Eqref{E7}. 
In the bottom plot we have $\vec{s}=-\vec{a}_4$ and \refeq{E8} applies.   

By comparison, for the standard six-output ellipsometer with counts
$n_{\xi\pm}$ of the respective detectors, the ML estimator is
\begin{equation}
  \label{eq:E9}
  \vec{S}=\biggl(\frac{n_{x+}-n_{x-}}{n_{x+}+n_{x-}},
                \frac{n_{y+}-n_{y-}}{n_{y+}+n_{y-}},
                \frac{n_{z+}-n_{z-}}{n_{z+}+n_{z-}}\biggr)\,,
\end{equation}
provided that the length of this $\vec{S}$ does not exceed unity.
If it does, the search has to be constrained to unit vectors, much like in the
discussion of Eqs.~\refeq{D6}--\refeq{D14}.
When \Eqref{E9} applies, the error corresponding to \refeq{E7} is given by
\begin{equation}
  \label{eq:E10}
   (\Delta\vec{s})^2 =\frac{9-3s^2}{N}\,.
\end{equation}
For $s>0$, this is a bit smaller than \refeq{E7}, so that the six-output
scheme gives a slightly more reliable estimate.
But when $s$ gets close to unity, the four-output scheme can provide better
estimates because there are states for which the error is substantially
smaller, as the extreme situation of \Eqref{E8} shows. 
For the six-output ellipsometer, the privileged pure states $\rho$ project on
an eigenstate of $\sigma_x$, or $\sigma_y$, or $\sigma_z$, so that one of the
six detectors in Fig.~\ref{fig:standard} never registers a qubit. 
Then we have
\begin{equation}
  \label{eq:E11}
   (\Delta\vec{s})^2 =\frac{8}{3N}\,,
\end{equation}
which is a bit larger than its analog \refeq{E8}.
The main lesson is thus that the two schemes are of comparable asymptotic 
efficiency.

\section{Optimality of minimal four-state tomography}\label{sec:optimal}
\subsection{Cram\'er--Rao bound}\label{sec:C-R}
As a preparation of the discussion below on the optimality and efficiency of
the minimal four-state tomography, we recall some well known facts 
about state estimation.
We regard the qubit state as parameterized by the three components of its
Pauli vector $\vec{s}=(s_x,s_y,s_z)$, rather than by the four probabilities 
$p_j$, because the latter are constrained. 
Then, the error of the components of the ML estimator $\vec{S}$ can be
estimated by the Cram\'er--Rao lower bound \cite{rao,cramer},
\begin{equation}\label{eq:F1}
(\Delta s_\xi)^2\ge \bigl(I_\mathrm{F}^{-1}\bigr)_{\xi\xi}
\qquad\mbox{with $\xi=x,y,z$,}
\end{equation}
where $I_\mathrm{F}$ is the Fisher information matrix,
\begin{equation}\label{eq:F2}
\bigl(I_\mathrm{F}\bigr)_{\xi\zeta}=
\overline{ 
\frac{\partial\log \mathcal{L}}{\partial s_\xi}
\frac{\partial\log \mathcal{L}}{\partial s_\zeta}
}\,,
\end{equation}
and the averaging is done over the data, that is over the multinomial
distribution that we exploited already in Sec.~\ref{sec:ManyClicks}.

There is no unique numerical measure for the comparison of the estimated state
$\rho_\mathrm{est}=\frac{1}{2}(1+\vec{S}\cdot\vec{\sigma})$ with the true
state $\rho_\mathrm{true}=\frac{1}{2}(1+\vec{s}\cdot\vec{\sigma})$. 
One can make a case for the trace-class distance
\begin{equation}
  \label{eq:F2a}
\mathcal{D}_\mathrm{tr}=\frac{1}{2}\tr{|\rho_\mathrm{est}-\rho_\mathrm{true}|}
=\frac{1}{2}\bigl|\vec{S}-\vec{s}\bigr|\,,  
\end{equation}
for the Hilbert--Schmidt distance 
\begin{equation}
  \label{eq:F2b}
\mathcal{D}_\mathrm{HS}
=\frac{1}{2}\bigl[\tr{(\rho_\mathrm{est}-\rho_\mathrm{true})^2}\bigr]^{1/2}
=\frac{1}{2}\bigl|\vec{S}-\vec{s}\bigr|\,,  
\end{equation}
for the Uhlmann fidelity 
\begin{eqnarray}
  \label{eq:F2c}
\mathcal{U}&=&\tr{|\sqrt{\rho_\mathrm{est}}\sqrt{\rho_\mathrm{true}}|}
\\\nonumber
&=&\frac{1}{2}\Bigl(1+\vec{S}\cdot\vec{s}+\sqrt{\cdot/\cdot}\Bigr)^{1/2}
+\frac{1}{2}\Bigl(1+\vec{S}\cdot\vec{s}-\sqrt{\cdot/\cdot}\Bigr)^{1/2}
\end{eqnarray}
with
\begin{eqnarray}
  \label{eq:F2d}
 \sqrt{\cdot/\cdot}=\sqrt{(\vec{S}+\vec{s})^2-(\vec{S}\times\vec{s})^2} 
\,,
\end{eqnarray}
and for some more.
In the single-qubit situation of present interest, there is no difference
between $\mathcal{D}_\mathrm{tr}$ and $\mathcal{D}_\mathrm{HS}$, but they are
not the same for higher-dimensional systems, such as the qubit pairs of
Sec.~\ref{sec:2qubits}. 
We note in passing that the Uhlmann fidelity and the trace-class distance are
natural pairs, inasmuch as they obey a fundamental inequality,
\begin{equation}
  \label{eq:F2e}
  \mathcal{D}_\mathrm{tr}^2+\mathcal{U}^2\leq1\,,
\end{equation}
irrespective of the dimension of the Hilbert space.

As long as one is comparing pure states with each other, the actual choice
between the quantitative measures of Eqs.~\refeq{F2a}--\refeq{F2c}
does not matter much, because then all of them are monotonic functions 
of the length of the difference $\vec{S}-\vec{s}$ 
of the two Pauli vectors.
If one state is mixed, however, this is not true for the Uhlmann fidelity. 
For the sake of computational simplicity, we opt for the (square of twice)
the Hilbert--Schmidt distance, and thus quantify the measure of the estimate
by the size of $(\vec{S}-\vec{s})^2$,
whose statistical average is already considered in Sec.~\ref{sec:asymptotics}.
A convenient conservative estimate is given by the Cram\'er--Rao bound 
of \Eqref{F1}.

According to Fisher's theorem \cite{Fisher} the ML estimator of $\rho$ 
attains the Cram\'er--Rao bound  for large $N$. 
Strictly speaking, this statement applies only to mixed states, 
because only for them the ML estimators are unbiased. 
For pure states, the positivity constraint plays an important role, and 
the Cram\'er--Rao bound derived for unbiased estimators tends to
overestimate the error, and therefore we shall consider pure states 
separately below.
As a rule of thumb, the Cram\'er-Rao bound 
is reliable when the lion's share of the likelihood function 
is contained within the Bloch ball.
This can, in fact, be used as an operational definition of ``mixedness''.
Indeed, as the analysis in Sec.~\ref{sec:ManyClicks} shows, for mixed states
there is no significant fraction of the likelihood function outside the Bloch 
ball if $N$ is sufficiently large.

\subsection{Optimality of the tetrahedron geometry}\label{sec:tetra-opt}
The four-state tomography with the tetrahedron geometry of
Eqs.~(\ref{eq:B1}--\ref{eq:B5}) is the best minimal qubit state 
measurement, in the sense that, among the measurements with 
four output channels, it provides the greatest  accuracy.
This can be seen as follows.
 
For the multinomial statistics of Sec.~\ref{sec:ManyClicks},
the Fisher information simplifies to
\begin{equation}
\bigl(I_\mathrm{F}\bigr)_{\xi\zeta}=N\sum_j\frac{1}{p_j}
\frac{\partial p_j}{\partial s_\xi}
\frac{\partial p_j}{\partial s_\zeta}\,,
\end{equation}
where $p_j$ is the probabilities of detecting a qubit in the $j$-th 
output channel. 
Consider now this variational problem:
For a given input state, the average distance $D$ is to be minimized
over all possible four-element POVMs.
The functional in question is 
\begin{equation}\label{eq:F4}
D=\mathrm{Sp}\bigl(I_\mathrm{F}^{-1}\bigr)-\lambda \sum_j \Pi_j\,,
\end{equation}
where $\mathrm{Sp}(\ )$ is the trace (``spur'') of the $3\times3$ matrix, 
and a Lagrange operator $\Lambda$ takes care of the constraint 
$\sum_j \Pi_j=1$. 

First, let us find the POVM that minimizes the 
functional \refeq{F4} for the maximally mixed input state. 
In that case the extremal equations read
\begin{eqnarray}\label{eq:F5}
R_j\Pi_j&=&\Lambda \Pi_j\qquad\mbox{for $j=1,\ldots,4$},
\nonumber\\
\Lambda&=&\sum_jR_j\Pi_j,
\end{eqnarray}
where
\begin{equation}\label{eq:F6}
R_j=\textrm{Sp}\bigl(I_\mathrm{F}^{-2} T_j\bigr)
\end{equation}
involves the operator matrix $T_j$ whose matrix elements are given by
\begin{equation}\label{eq:F7}
T_{j,\xi\zeta}=\tr{\Pi_j\sigma_\xi}\sigma_\zeta+
\sigma_\xi\tr{\sigma_\zeta\Pi_j}\,.
\end{equation}
Note the joint appearance of $3\times3$ $\mathrm{Sp}(\ )$ traces and 
quantum-mechanical $\tr{\ }$ traces. 
One verifies by inspection that any POVM of the tetrahedron geometry of
Eqs.~(\ref{eq:B1}--\ref{eq:B5}) satisfies these extremal equations
and hence identifies the maximally mixed state with greatest accuracy.

Obviously, for biased states the optimal POVM itself becomes biased.
The generalization of the extremal equations to this case is 
straightforward; the operators $T_j$ will then contain one more term
proportional to the true state $\rho$. 
Although an analytical solution may be difficult to obtain, one can always 
find the extremal measurement by an iterative procedure. 
As the input state is usually not known
and selected in random, operators $R_j$ should be averaged over the
Bloch ball. In this way one obtains an algorithm providing
the optimal sequential measurement, in the sense that the
Hilbert-Schmidt distance is consistently reduced in each iteration step. 
Numerical results show that the tetrahedron 
POVM is also optimal for uniformly distributed input states.

Quite explicitly, the optimal distance for the tetrahedron POVM is
\begin{equation}\label{eq:F8}
D_\textrm{opt}=\frac{9-s^2}{N}\,,
\end{equation}
in agreement with, or as a consequence of, the mean square distance 
of \Eqref{E7}.
As one would expect, the accuracy of the ellipsometer is somewhat better
for pure input states than for mixed states.
But, most importantly, the accuracy 
does not depend on the orientation of $\vec{s}$ relative to the measurement 
tetrahedron.

\subsection{Orientation of the measurement tetrahedron}\label{sec:orient}
Their relative orientation is, however, not completely irrelevant.
For example, the value of $\kappa^2$ in \refeq{E5} clearly depends on it.
More generally, we can obtain the large-$N$ mean values of functions of 
$\vec{S}-\vec{s}$ from the asymptotic generating function
\begin{equation}
  \label{eq:F9}
  \overline{\exp\Bigl(i\vec{r}\cdot\bigl(\vec{S}-\vec{s}\,\bigr)\Bigr)}
=\exp\Bigl(-\frac{1}{2N}\vec{r}\cdot\tensor{K}\cdot\vec{r}\Bigr)\,,
\end{equation}
where the dyadic
\begin{equation}
  \label{eq:F10}
  \tensor{K}=9\sum_{j,k}\vec{a}_j(\delta_{jk}p_j-p_jp_k)\vec{a}_k
\end{equation}
depends on the positioning of $\vec{s}$ relative to the vector quartet of
the $\vec{a}_j$'s.
This is, of course, an immediate consequence of \Eqref{E6}.

As an application, let us consider the asymptotic mean value of the Uhlmann 
fidelity of Eqs.~\refeq{F2c} and \refeq{F2d}.
We first note that
\begin{equation}
  \label{eq:F11}
  \mathcal{U}=1-\frac{1}{8}\left[
\bigl(\vec{S}-\vec{s}\,\bigr)^2+
\frac{\Bigl(\vec{s}\cdot\bigl(\vec{S}-\vec{s}\,\bigr)\Bigr)^2}{1-s^2}\right]
\end{equation}
holds when $\vec{S}-\vec{s}$ is small.
Then we recall \Eqref{E7} and extract
\begin{equation}
  \label{eq:F12}
  \overline{\Bigl(\vec{s}\cdot\bigl(\vec{S}-\vec{s}\,\bigr)\Bigr)^2}=
\frac{1}{N}\vec{s}\cdot\tensor{K}\cdot\vec{s}=\frac{72\kappa^2}{N}
\end{equation}
from \refeq{E9} to arrive at
\begin{equation}
  \label{eq:F13}
  \overline{\,\mathcal{U}\,}=1-\frac{9-s^2}{8N}-\frac{9\kappa^2}{N(1-s^2)}\,,
\end{equation}
where we meet the orientation-dependent quantity  $\kappa^2$ of \Eqref{E5}.

There are extremal orientations of three kinds.
The value of $\kappa^2$ is largest when $\vec{s}$ is parallel to one of the
vectors of the tetrahedron quartet, and smallest when $\vec{s}$ is
antiparallel. 
In addition to these four maxima and four minima of $\kappa^2$, there are also
six saddle points that have $\vec{s}$ parallel to the sum of two different
vectors of the quartet.
The maximal and minimal values of $\kappa^2$ are given by the upper and lower
signs in 
\begin{equation}
  \label{eq:F14}
  \kappa^2=\frac{(1\pm s)(3\mp s)s^2}{72}\quad
\mbox{for $\vec{s}=\pm s\vec{a}_j$}\quad\mbox{(any $j$),}
\end{equation}
respectively, and the value at the saddle points is
\begin{equation}
  \label{eq:F15}
  \kappa^2=\frac{(3-s^2)s^2}{72}\quad
\mbox{for $\vec{s}=\displaystyle\frac{\sqrt{3}}{2}\,s
\bigl(\vec{a}_j+\vec{a}_k\bigr)$}\quad\mbox{(any $j\neq k$).}
\end{equation}
Accordingly, the approach of $\overline{\,\mathcal{U}\,}$ to unity is fastest 
for the $\vec{s}=-s\vec{a}_j$ orientation, for which the large-$N$
approximation 
\begin{equation}
  \label{eq:F16}
  \overline{\,\mathcal{U}\,}=1-\frac{(3+s)(3+2s)}{8N(1+s)}
\end{equation}
applies.
A small value of $\kappa$ is also advantageous in \Eqref{E4}, as it ensures
a large negative argument of the error function and thus a small probability
for violating the inequality \refeq{D5}.

\section{Measuring pure qubit states}\label{sec:pure}
When the measured quantum system is known to be in a  pure state
--- which is a bold over-idealization of any realistic situation --- 
this knowledge can be exploited systematically when estimating the otherwise
unknown state.
A somewhat more realistic situation arises when the input state is pure
but we do not know this to begin with, although this case is quite a bit
artificial as well, because one can hardly assume that real sources are 
not affected by classical noise, or that the experimental setup is totally 
decoupled from the environment.
Put differently, it is far-fetching to assume that the experimenter will ever
have the perfect control that is necessary to ensure that a source emits a
pure state. 
Nevertheless, there is an interest in such idealized scenarios because they
come up in analyses of eavesdropping attacks on schemes for quantum
cryptography, where --- as a matter of principle --- it is assumed that the
eavesdropper is only limited by the laws of physics, not by practical
limitations. 

For $s=1$, the Uhlmann fidelity \refeq{F2c} simplifies to
\begin{equation}
  \label{eq:G0a}
  \mathcal{U}=\sqrt{\frac{1+\vec{s}\cdot\vec{S}}{2}}\,,
\end{equation}
and is equivalent to
\begin{equation}
  \label{eq:G0b}
   \mathcal{U}=\sqrt{1-\frac{1}{4}\bigl(\vec{S}-\vec{s}\bigr)^2}
\end{equation}
if the estimator is a pure state as well, $S=1$.
When indeed estimating pure states with pure states, the situation to be
considered now, it is customary to take the average of $\mathcal{U}^2$,
\begin{equation}
  \label{eq:G1}
  \overline{F}=\overline{\mathcal{U}^2}
=1-\frac{1}{4}\overline{\bigl(\vec{s}-\vec{S}\bigr)^2}
=1-\frac{1}{4}(\Delta\vec{s})^2\,,
\end{equation}
as the fidelity measure that judges the quality of the estimation procedure.
One must keep in mind that both the true Pauli
vector $\vec{s}$ and its estimator $\vec{S}$ are unit vectors here, as
\Eqref{G1} applies only under this restriction.
Knowing that the source generates pure states means having
a lot of prior knowledge because the set of pure states is much smaller
than the set of mixed states (the Bloch sphere rather than the Bloch ball),
and so one can safely expect that a better, possibly much better, accuracy 
of the estimation can be achieved. 

The average fidelity of the optimal \emph{joint} measurement on $N$ copies is 
known to obey the inequality \cite{massar-popescu,derka-buzek}
\begin{equation}\label{eq:G2}
\overline{F}\leq\frac{N+1}{N+2},
\end{equation}
so that the corresponding error $1-\overline{F}$ will decrease 
as  $1/N$ in the large-$N$ limit. 
Any reasonable estimator should show this dependence.

In our scheme the qubits are measured individually, not jointly.
Nevertheless, 
as a consequence of the Fisher theorem in general, and of the findings in
Sec.~\ref{sec:asymptotics} in particular, the variance $(\Delta\vec{s})^2$ in
\Eqref{G1} is proportional to $1/N$ in the large-$N$ limit, and so is then
$1-\overline{F}$. 
But we need to reconsider the asymptotics of the maximum-likelihood
estimation, now taking into account that both $\vec{s}$ and $\vec{S}$ are
restricted to the Bloch sphere, so that 
we get the estimator $\vec{S}$ from Eqs.~\refeq{D10}--\refeq{D13}, whether
inequality \refeq{D5} is obeyed or not.

Here, too, the estimation is sensitive to the orientation of the Pauli vector
$\vec{s}$ relative to the vector quartet of the measurement tetrahedron.
We deal first with the case of ``not anti-aligned,'' that is
$\vec{s}\neq-\vec{a}_j$ for all $\vec{a}_j$'s.   
Then, $\nu_j-p_j$ and $\mu$ are of the order of $1/\sqrt{N}$ and $\mu$ is given
by
\begin{equation}
  \label{eq:G3a}
  \mu=\frac{2}{3\kappa^2}\sum_jp_j(\nu_j-p_j)
\end{equation}
with $\kappa^2>0$ from \Eqref{E5}.
The resulting averages
\begin{eqnarray}
  \label{eq:G3b}
  \overline{(\nu_j-p_j)\mu}&=&\frac{2}{9N\kappa^2}(3p_j-1)p_j\,,
\nonumber\\
\overline{\mu^2}&=&\frac{2}{9N\kappa^2}\,,
\end{eqnarray}
available as a consequence of \Eqref{E2},
are used in 
\begin{equation}
  \label{eq:G3c}
  \overline{F}=1-3\sum_j\overline{\bigl[(\nu_j-p_j)-(3p_j-1)p_j\mu\bigr]^2}
\end{equation}
to arrive at
\begin{eqnarray}
 \label{eq:G3d}
1-\overline{F}&=&
\frac{2}{N}-\frac{2}{3N\kappa^2}\sum_j\bigl[(3p_j-1)p_j\bigr]^2
\nonumber\\
   &=&\frac{4}{N}-\frac{2}{9N\kappa^2}\Bigl(27\sum_jp_j^4-1\Bigr)\,.
\end{eqnarray}

In particular, when the tetrahedron is aligned with the Pauli vector,
the upper-sign case of \refeq{F14} with $s=1$, we have $\kappa^2=1/18$ and
$27\sum_jp_j^4=7/4$, so that
\begin{eqnarray}\label{eq:G4}
1-\overline{F_{\uparrow\!\uparrow}}=\frac{1}{N}
\end{eqnarray} 
for this parallel strategy ($\uparrow\!\uparrow$).
This is only half as big as what one would get for the $s=1$ value of 
\Eqref{E7}, and thus demonstrates the advantage of estimating the pure true
state by pure-state estimators only.

There is no such advantage for the anti-parallel strategy 
($\uparrow\downarrow$) that has the tetrahedron anti-aligned with $\vec{s}$,
because the argument in Sec.~\ref{sec:ManyClicks} establishes that the ML
estimator is always a pure state then.
Since this is now the lower-sign case of \refeq{F14} for $s=1$, we have
$\kappa^2=0$ and Eqs.~\refeq{G3a}--\refeq{G3d} are not valid.
Instead, we recall that Eqs.~\refeq{E8a}--\refeq{E8} apply, and conclude that
\begin{equation}
  \label{eq:G5}
  1-\overline{F_{\uparrow\!\downarrow}}=\frac{1}{2N}
\end{equation}
holds here. 
The anti-parallel strategy has thus half the average error of the aligned
strategy. 

On the other hand, for a generic orientation of the input state
there will always be an uncertainty in the length of the input state
vector and therefore the fidelity will approach unity at a much 
slower rate proportional to $1/\sqrt{N}$.
Only if one knows beforehand that the input state is pure,
the $1/N$ rate can be achieved even for general orientations
of the tetrahedron. 
Let us illustrate this last remark by the example of the parallel strategy.
For the parallel orientation, only about half of the measurement outcomes
will violate inequality \refeq{D5}, the others would not,
see Sec.~\ref{sec:ManyClicks}.
Knowing that the input state is pure one can, however, ignore this inequality
and always solve Eqs.~\refeq{D8}--\refeq{D10}
for the input state unit Pauli vector.

\section{Adaptive strategies}\label{sec:adaptive}
\subsection{A pre-measurement strategy}\label{sec:premeasure}
The good performance of the anti-parallel strategy hints at a very simple 
adaptive procedure that provides the fast
$1/N$ asymptotic behavior without any prior knowledge about the purity
of the input state: Let us split the input ensemble into two
halves. After the first $N/2$ particles have been registered
and the direction of the input Pauli vector estimated,
the experimenter adopts the anti-parallel strategy for measuring
the rest of the ensemble.
Notice that in this simple adaptive scenario, the first half of the particles 
are used for a pre-measurement and serve only for adjusting the measurement 
apparatus, while it is the second half which provide the actual estimate 
of the input state.

Let $\theta$ denote the angle between the Pauli vector $\vec{s}$ 
of the input state and the Pauli vector $\vec{s}_1$ estimated from the first 
half of the ensemble.
This angle $\theta$ can be estimated with an accuracy of 
$1-\overline{\cos\theta}\propto 1/N$ in the first stage. 
This means that, in the second stage, the mean probability 
of detecting a particle in the channel anti-parallel to 
$\vec{s}_1$ will be proportional to $1/N$.
No matter how large is $N$, only a few particles will be detected
in this channel. 

The maximal uncertainty in the length of the Pauli vector
is then easily calculated with the aid of 
\begin{equation}\label{eq:H1}
S^2
=\frac{12}{N^2}\sum_j n_j^2-3
\end{equation}
for the estimator of \Eqref{D2}.
Let us set $n_1=\delta$ with $\delta$ a small number independent of $N$,
and look for the minimal length of the estimated Pauli vector compatible 
with the given $\delta$. Since the right-hand side of \Eqref{H1} is a 
concave function of $n_2$, $n_3$, and $n_4$, it is minimal
when all of them are equal to each other, 
$n_2=n_3=n_4=(N-\delta)/3$, with the consequence
\begin{equation}\label{eq:H2}
S_\mathrm{min}=1-\frac{4\delta}{N}
\end{equation}
for large $N$.
This guarantees that in the second stage both the orientation
and length of the Pauli vector, and so the fidelity, will be determined 
with an error proportional to $1/N$.  
Numerical simulations show that for large $N$ (up to $N=10^5$)
the mean error $1-\overline{F}$ of this simple protocol is about twice 
as large as, and thus worse than, that of the optimal joint POVM measurement,
in which all $N$ qubits are measured together.
Let us emphasize that while our protocol would provide this performance for
any input pure state, it also provides a meaningful estimate
for any input mixed state, of course, with a larger uncertainty proportional to
$1/\sqrt{N}$. In contrast to that, the optimal joint measurement
that attains the ultimate limit of \Eqref{G2} 
would not work for mixed input states.

\subsection{Self-learning strategies}\label{sec:SelfLearn}
It is known that sequential measurements when combined with self-learning 
adaptive strategies can come close to the quantum estimation limit 
\cite{freyberger-pure,freyberger-mixed,adaptive-experiment}.
Their improvement on the conventional sequential measurement depends
on the purity of the input state \cite{freyberger-mixed}.
Adaptive techniques are more sensitive to pure states than to 
states with a lot of classical noise.

The optical network of Fig.~\ref{fig:minimal}(b) 
can easily be adapted to a self-learning procedure, 
and so can other optical implementations. 
After each detection the current information about
the input state can be evaluated, and the operations $\sigma_x$, $\sigma_y$, 
and $\sigma_z$ acting on the next particle inside the interferometer
can be modified by a common unitary transformation.
This is economically achieved by performing the required unitary
transformation on the approaching photon before it enters the interferometer
proper.  

From the discussion in Sec.~\ref{sec:pure}
one might get the impression that a particularly good adaptive strategy 
would be to always keep one of the measured half-projectors anti-parallel
to the current estimate. 
Matters are, however, not so simple.

Let us illustrate the difference between the adaptive and the non-adaptive
procedure at the extreme example of measuring only two qubits. 
In the \emph{non-adaptive} case, everything is as discussed above, 
in particular the probability that the first qubit is detected by the 
$j$-th detector and the second by the $k$-th detector is 
$p_jp_k$ with the $p_j$'s of \Eqref{B6} and, in accordance with
Sec.~\ref{sec:Counting}, the ML estimator is $\vec{S}=\vec{a}_j$ if $j=k$ and
$\vec{S}=\sqrt{3/4}(\vec{a}_j+\vec{a}_k)$ if $j\neq k$.
Upon averaging $(\vec{S}-\vec{s})^2$ first over all measurement results for 
a given input Pauli vector $\vec{s}$ and then over all possible inputs, we
thus get
\begin{equation}
  \label{eq:H3}
  \overline{\bigl(\vec{S}-\vec{s}\bigr)^2}=\frac{5-\sqrt{3}}{3}=1.089
\end{equation}
as the figure of merit.

In the \emph{adaptive} case, the tetrahedron is realigned for the second qubit
after the first qubit has been detected.
Since the ML estimator $\vec{S}$ obtained after the detection 
of the first qubit
will coincide with one of the tetrahedron vectors, the anti-aligning of the
tetrahedron for the second qubit amounts to the replacement 
$\vec{a}_j\to-\vec{a}_j$. 
Now the probability that the first qubit is detected by the 
$j$-th detector and the second by the $k$-th detector is 
$\mathcal{L}=p_j(\frac{1}{2}- p_k)$.
The resulting ML estimator is then given by 
$\vec{S}=\sqrt{3/8}(\vec{a}_j-\vec{a}_k)$, and upon averaging over all
measurement results and all input Pauli vectors we get
\begin{equation}
  \label{eq:H4}
  \overline{\bigl(\vec{S}-\vec{s}\bigr)^2}=\frac{11-\sqrt{24}}{6}=1.017\,,
\end{equation}
which is markedly smaller than the non-adaptive value in \refeq{H3}.
It is also smaller that the value for the adaptive strategy with parallel
alignment because that is identical with the non-adaptive procedure when only
two qubits are detected.

We note that the averages in \refeq{H3} and \refeq{H4} are taken over 
pure input states.
If one averages over all input states, pure and mixed,  
one obtains the respective numbers $(7-\sqrt{3})/5=1.054$ and
$(7-\sqrt{6})/5=0.910$,
again with a clear advantage for the anti-aligning adaptive strategy.

\begin{figure}
\centerline{
\includegraphics{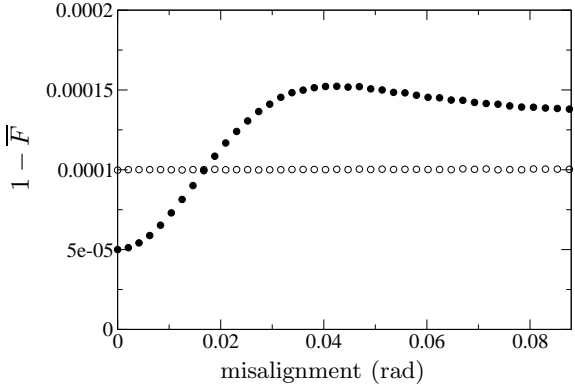}
}
\caption{\label{fig:sens}Mean estimation errors
of the POVM tetrahedrons that, on the Bloch sphere, differ from the
exact parallel (squares)  and anti-parallel (circles) orientations by
the known angle given on the abscissa. }
\end{figure}

\subsection{Numerical simulations}\label{sec:NumSim}
Although these numbers speak clearly in favor of the anti-aligning adaptive 
scheme, one should, however, keep in mind that they apply only
for the exactly aligned or anti-aligned settings of the apparatus.
But, in such an adaptive scheme, after the first particle is observed,
the uncertainty of the input state Pauli vector orientation is 
still quite large, which may result in a significant misalignment
in the second adaptive step.
In fact, the anti-aligning strategy has a greater sensitivity 
to such misalignments. 
This is illustrated by the simulation data shown in Fig.~\ref{fig:sens}, 
where the estimation errors of both strategies are shown in dependence on 
the misalignments of the 
apparatuses for the chosen input intensity of $10^4$ particles. 
Each point has been obtained by averaging over $5\times10^5$ ML estimates.

As expected, the anti-aligning adaptive  strategy
performs better if no misalignment is present.
However, even a small misalignment (of the order of $1^\circ$ 
in this case) is enough
to wash out this advantage. For even larger misalignments
the aligned setting provides much better performance \cite{BayesAver3}.
Different sensitivities of both measurement strategies to this
kind of error might be of quite some importance for the potential 
applications of the minimal ellipsometer in quantum communication 
protocols and quantum cryptography.

\begin{figure}
\centerline{\includegraphics{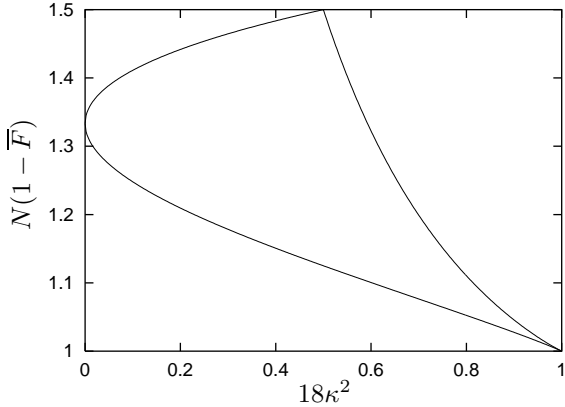}}
\caption{\label{fig:pure}%
For each value of $\kappa^2$ between $\kappa^2=0$ (perfect anti-alignment) and
$\kappa^2=1/18$ (perfect alignment), the possible values of the coefficient 
of the $1/N$ term in \Eqref{G3d} are in the area bounded by the two curves.
The smallest coefficient obtains for the case of perfect alignment, when
\Eqref{G4} applies.
}
\end{figure}

One can understand this extreme sensitivity of the anti-aligned setting,
and why it becomes immediately worse than the aligned setting, 
by a second look at Eqs.~\refeq{G3d} and \refeq{G5}.
The $\kappa^2=0$ result \refeq{G5} is \emph{not} the $\kappa^2\to0$ limit of
the $\kappa^2>0$ result \refeq{G3d}. 
In fact we have, see Fig.~\ref{fig:pure},
\begin{equation}
  \label{eq:H5}
  1-\overline{F}\to \frac{4}{3N}\quad\mbox{as $\kappa^2\to0$ in \Eqref{G3d},}
\end{equation}
which is larger than the error of $\overline{F_{\uparrow\!\uparrow}}$ in
\Eqref{G4}.  
Therefore, the slightest misalignment takes us from the $1/(2N)$ error of
\Eqref{G5} to this $4/(3N)$ because \refeq{H5} applies to tiny non-zero
values of $\kappa^2$, while \refeq{G5} holds only if $\kappa^2=0$ exactly.

This observation also resolves the apparent contradiction between the general
upper bound of \Eqref{G2} and the large-$N$ error for perfect anti-alignment 
in \Eqref{G5}, which does not respect that upper bound.
Nevertheless, this example is not a valid counterexample, because it refers to
an absurdly artificial situation: The experimenter has perfect a priori
knowledge of the state to be measured and has perfect control over his
measurement apparatus, such as to ensure the perfect anti-alignment to which
\Eqref{G5} applies. 
In other words, \emph{when} it applies, there is no need for a state
estimation to begin with.

\begin{figure}[b]
\centerline{
\includegraphics{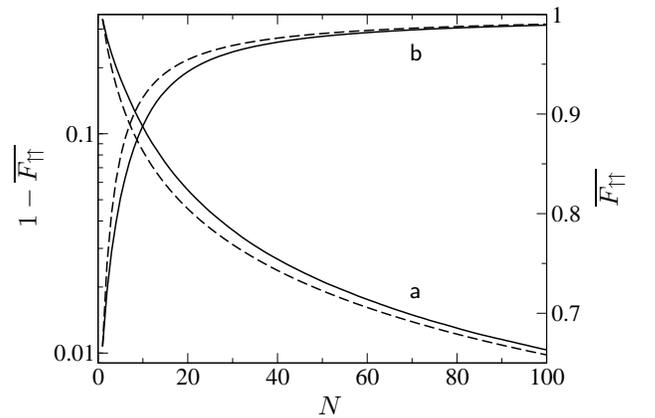}
}
\caption{\label{fig:fidel}%
Mean error (curves \textsf{a}, decreasing) and mean fidelity 
(curves \textsf{b}, increasing) of the minimal qubit tomography as a function
of $N$, the size of the measured ensemble. 
Solid lines: the proposed network of Figs.~\ref{fig:minimal} and 
\ref{fig:QCnet} with the adaptive parallel strategy described in the text; 
broken lines: the quantum limit of \Eqref{G2}. 
Notice that the proposed optical network attains the quantum limit 
asymptotically for large $N$.
}
\end{figure}

Having thus compared the performances of the two extreme strategies,
we now calculate the mean fidelity of the 
following adaptive measurement: After detecting each new qubit 
the information about the input state is updated and a new
ML estimate is calculated.
Then one of the measured half-projectors is aligned along this
current estimate.
These two steps are repeated until all input particles are 
used up. Figure \ref{fig:fidel} shows mean 
fidelities and errors that were obtained by averaging over $200'000$ 
randomly selected pure input states. The quantum limit, 
\Eqref{G2}, is shown for comparison. 
It is evident that this bound can be attained only for 
large $N$, while the most pronounced difference is seen for 
moderately-sized ensembles.
Such a behavior is typical for all sequential
self-learning estimation strategies.

\begin{figure}[t]
\centerline{
\includegraphics{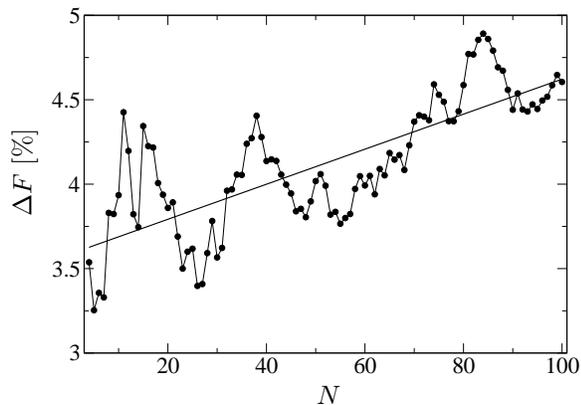}
}
\caption{\label{fig:random}%
Relative difference $\Delta F$, in percent, between the errors of the
random and parallel adaptive strategies (black dots); least-square linear
fit (straight line). 
The lines connecting the dots guide the eye but have no further significance.
The irregularities stem from the limited number of input states 
(200'000) used for the averaging.
}
\end{figure}

Finally, let us compare the efficiency of the parallel 
adaptive strategy with a very simple sequential measurement 
where the orientation of the measured half-projectors is chosen
at random in each step, see Fig.~\ref{fig:random}.
As expected, the adaptive strategy is better and its benefit 
grows with increasing size of the measured ensemble.

\section{Summary}\label{sec:sum}
We have presented a minimal measurement scheme for single-qubit tomography
that has no more than the necessary number of four outputs.
The scheme is conceptually simple, highly symmetric and optimal among all
four-output schemes, and can be realized with the present technology for the
polarization qubit of photons emitted by a single-photon source. 
As a demonstration, we designed a simple, but not simplest, optical network.

Our thorough analysis showed that the scheme is efficient in the sense that it
enables one to estimate the qubit state reliably without first detecting an
enormous number of qubits --- a few thousand are sufficient for most practical
applications, a few hundred may be enough if extreme precision is not
required. 
The efficiency can be increased by adaptive procedures in which the apparatus
is adjusted in accordance with the current estimate for the qubit state.

Since the four-output setup provides optimal
complete tomography with the minimal number of output channels,
it is particularly well suited as a detection device for certain 
quantum communication protocols such as tomographic quantum cryptography
\cite{tomocrypt}.
Indeed, there are protocols for quantum key distribution           
that exploit the tetrahedron quartet of states \cite{RenesTetra},  
among them a highly efficient tomographic protocol \cite{minQKD}.

\begin{acknowledgments}
We are very grateful for the valuable discussions with Artur Ekert,
Christian Kurtsiefer, Antia Lamas-Linares, Ng Hui Khoon, Tin Kah Ming, 
and Goh Choon Guan. 
J.~\v{R}. wishes to thank for the kind hospitality during his visits to
Singapore. 
We gratefully acknowledge the financial support from 
Temasek Grant WBS: R-144-000-071-305, from
NUS Grant WBS: R-144-000-089-112, and from Grant No.~LN00A015 
of the Czech Ministry of Education.
\end{acknowledgments}



\begin{thebibliography}{99}
\bibitem{multiport} 
It may be worth mentioning that, in principle if not in practice, any
single-photon POVM can be implemented with an optical multiport in a rather
transparent fashion. ---
The setup of Fig.~\ref{fig:minimal}(b) is a four-port:
the photodetectors register the photons exiting at the four output ports; 
one of the two entry ports at the ${4:2}$ beam splitter is used, 
whereas no use is made of the two other entry ports at the ${3:1}$ 
and ${1:1}$ beam splitters.
The setup of Ref.~\cite{RenesDiss} is also a four-port, whereas those
Refs.~\cite{1loop,newSetup} are two-ports.
\bibitem{BayesAver1}
For example, one could take the likelihood of \Eqref{D3} as the weight function
for the calculation of the mean of all Pauli vectors $\vec{s}$,
thus averaging over the part of the cloud in
Fig.~\ref{fig:clouds} that is inside the Bloch sphere, and then
accept this average as the best guess.
---
Consult the classic text by Helstrom \cite{Helstrom} for a general
introduction to quantum state estimation, or turn to Ref.~\cite{LNP} for a
very recent account.
\bibitem{BayesAver2}
By construction, the procedure of \cite{BayesAver1} would always give a
physically acceptable estimate.
\bibitem{BayesAver3}
Further support in favor of the aligned adaptive scheme is given by the
figures of merit obtained for the two-qubits-only situation of
Sec.~\ref{sec:SelfLearn} by the procedure of \cite{BayesAver1}.
The numbers corresponding to those in Eqs.~\refeq{H3} and \refeq{H4}
are $11/21$ and $21/40$, when all input states are averaged over, as well as
$19/24$ and $4/5$, when the average is over pure states only.
In both cases, the aligned scheme does slightly better than the anti-aligned
scheme.
\end{thebibliography}
\end{document}